\documentstyle[epsf,12pt]{article}
\begin{document}
\begin{titlepage}
\begin{center}
\vspace*{3cm}

\begin{title}
\bold
{\Huge Entropy in cluster analysis \\of single events in heavy ion collisions

 }
\end{title}

\vspace{2cm}

\begin{author}
\Large
K. FIA{\L}KOWSKI\footnote{e-mail address: uffialko@thrisc.if.uj.edu.pl},
R. WIT\footnote{e-mail address: wit@thrisc.if.uj.edu.pl}

\end{author}

\vspace{1cm}

{\sl M. Smoluchowski Institute of Physics\\ Jagellonian University \\

30-059 Krak{\'o}w, ul.Reymonta 4, Poland}

\vspace{3cm}

\begin{abstract}
We analyse the cluster structure of the final multihadron states resulting 
from heavy ion collisions using the concept of Jaynes - Shannon entropy. 
Further evidence for an interesting differentiation of events is provided.
\end{abstract}

\end{center}

PACS:  25.70.-z\\

{\sl Keywords:} heavy ion collisions, cluster analysis, Monte Carlo  \\

\vspace{1cm}

\noindent

 4 October, 1999 \\

\end{titlepage}

\section{Introduction}
\par
The analysis of the KLM collaboration data~\cite{klm} for Pb(158 GeV/nucleon)-Ag/Br 
collisions indicated 
the existence of events with "strong dynamical fluctuations". The use of 
factorial moments as a tool for preliminary selection of events was suggested. In our 
previous 
paper ~\cite{fw} we analyzed the cluster structure of events resulting from the heavy 
ion collision data and from the Monte Carlo (MC) generators. We investigated the 
dependence of the cluster  distributions on the 
parameter defining a cluster size. The quantity we worked with was the second scaled 
factorial moment as a function of averaged multiplicity $\overline n$ with "two 
subtracted" (i.e. $\overline n$ - 2):

\begin{equation}
\tilde{F}_2 = \frac {\overline {(n-2)(n-3)}}{(\overline n -2)^2}.
\end{equation}

\par 
We gave also a motivation for the choice of such a discrimination tool. In full 
agreement 
with the suggestions following from Ref. ~\cite{klm} we have found two different patterns 
in the first four published experimental events. 
\par
In this note we would like to present some further analysis concerning the same subject.
We again  employ the
cluster definition used in the procedure implementing the Bose-Einstein
(BE) effect in MC ~\cite{fww} and in what follows only clusters containing at least two 
hadrons are taken into account.
\par 

As before, we work in a two-dimensional momentum space ($\eta -
\phi$) since in the data set ~\cite{klm} only the pseudorapidity $\eta$
and the azimuthal angle $\phi$ are given. As a distance measure for clustering we use

\begin{equation}
\delta^2 = (\Delta \eta)^2 + (\Delta \phi)^2
\end{equation}

\noindent
which provides also rather stable results under some coefficient change in front of
these two terms defining $\delta^2$.

\par In our procedure
originally each particle is considered as a single cluster. We fix the
value of a "cluster size parameter" $\epsilon $ (the upper limit of  $\delta^2$, for which
two particles are joined into one cluster) and perform the 
clustering
procedure for all pairs of particles.    
\par Our procedure is
different from the clustering procedure applied in ~\cite{klm}. Details are given in 
~\cite{fw}.  Here we would only like to point out that our results are significant if we
have a sufficiently large number of clusters (with at least two
particles). This limits the possible range of $\epsilon$ values, which
must neither be too small nor too large. Twenty values of $\epsilon$ ($\epsilon_i = 0.002*i$) 
were taken for each of the events under consideration.
\par 
Thus for each value of $\epsilon$ for a given event we have
\begin{itemize}
\item a number of particles in each cluster $n_k$
\item a number of clusters with $n_k > 1$, $N(\epsilon) \equiv N(i)$.
\end{itemize}
Obviously in the following $k = 1,...,N(\epsilon)$.
With this information we may also calculate for subsequent  values of $\epsilon$ the 
entropy ~\cite{jay}

\begin{equation}
S(\epsilon) = - \sum_k p_k  ln(p_k)
\end{equation}  

\noindent
where the summation runs over all clusters in an event (leaving aside one particle 
clusters). The quantity $p_k = n_k/\sum_k n_k$ is a  probability to find a particle in the 
k-th  cluster. 

As well known, the Jaynes - Shannon entropy introduced in this way is a good 
measure of the 
"amount of uncertainty" represented by a discrete probability distribution. 
Therefore one may expect that it will help (in a natural way) to distinguish between 
different heavy ion collision events. Here we do not use in our clustering procedure 
the principle of event entropy maximisation. A futher study of this interesting subject 
is in progress.  

\section{Analysis}
We performed the clustering procedure for
\begin{itemize}
\item  the "random events" of similar multiplicities obtained
by using a plain uniform random generator of $(\eta, \phi)$ points (SERENE events)
\item the four KLM events presented in ~\cite{klm} 
\item the events obtained from the VENUS generator ~\cite{ven}.
\end{itemize}
\par 
We have chosen randomly for presentation 5 out of 20 events generated for both the
 SERENE and VENUS classes.
The results for cluster distribution are presented in Figs. 1a)-3a).  
\par 
The number of clusters for all the SERENE events has a rather broad maximum for $i$ 
between 5 and 10 ($0.01 < \epsilon < 0.02$) with the value around 200. 
For VENUS events the maximum 
occurs at smaller $i$ and the subsequent decrease is much faster. 
Two of the experimental events resemble VENUS events, the other two are 
more similar to SERENE events. 
\par
The striking structure seen in Figs. 1a)-3a) appears again if we now plot (fot the  same 
range of $\epsilon$) the entropy defined in Eq. (3), as shown in Figs. 1b)-3b).
\par 
The maximum for all the SERENE events is very flat, whereas for all the VENUS 
events the values of entropy decrease rather fast for higher values of $\epsilon$.
The experimental events are again much more differentiated. More quantitative 
comparison of the events is possible if we define for each event 
\begin{itemize}
\item  the ratio $\rho_c$ of the maximal number of clusters in each event to the 
 number of clusters for the last 'bin' in $\epsilon$ in this event (cf. Figs. 1a)-3a))
\item the ratio $\rho_s$ of the maximal value of the entropy in each event 
to its "last bin" value (cf. Figs. 1b)-3b)).
\end{itemize}
\par
The values of these ratios are given in Tab.1. We checked that omitting clusters 
with only two particles we get similar results as shown in Tab.2, although the number 
of clusters and the values of entropy (especially for small values of $\epsilon$) 
are strongly reduced.
\par
The values 
of considered ratios for all SERENE events are very similar. For VENUS events
the values are much bigger and more differentiated. For two experimental events we 
find the values similar to those of  SERENE events and for two others the values  
are more similar to VENUS events. 

\begin{figure}
\centerline{%
\epsfxsize=7cm
\epsfbox{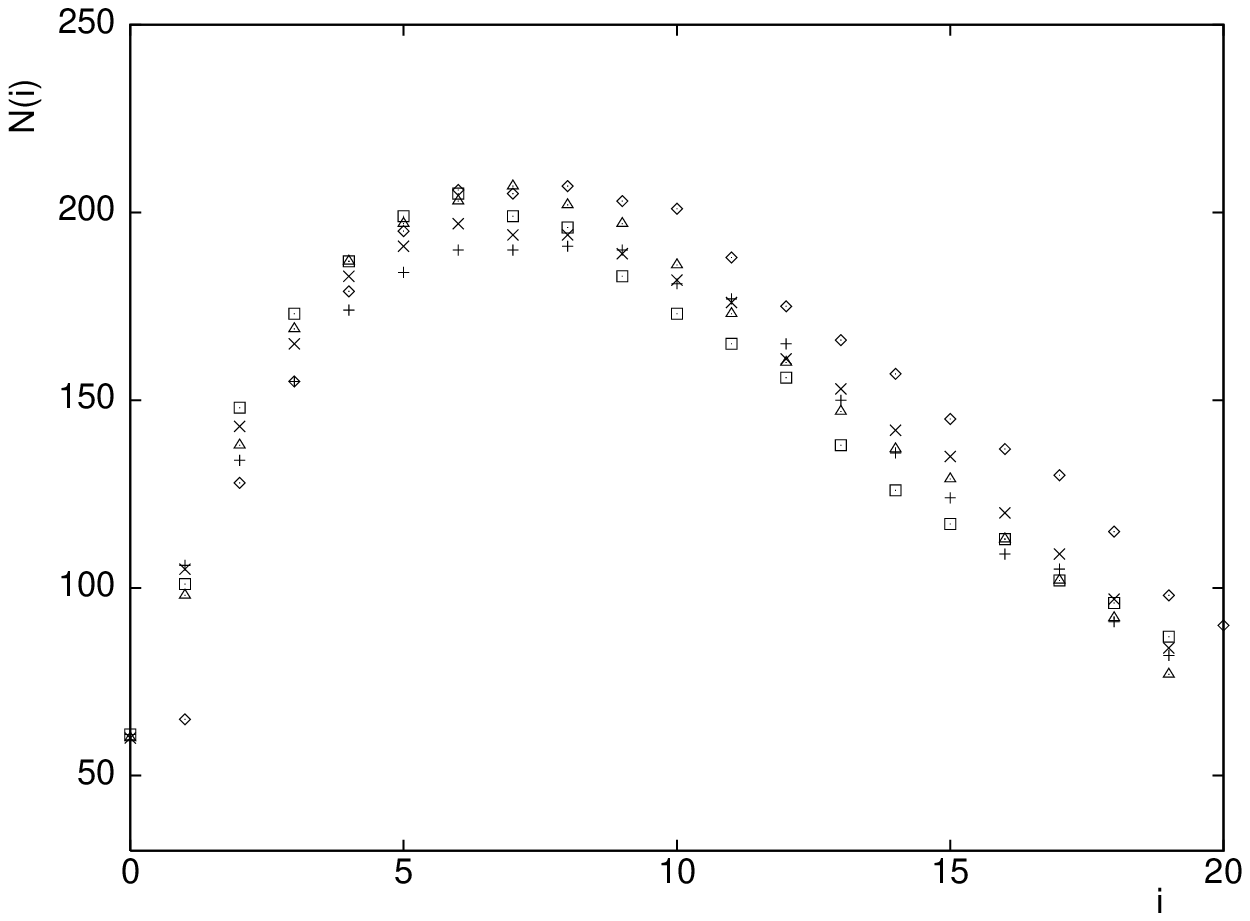}
\epsfxsize=7cm
\epsfbox{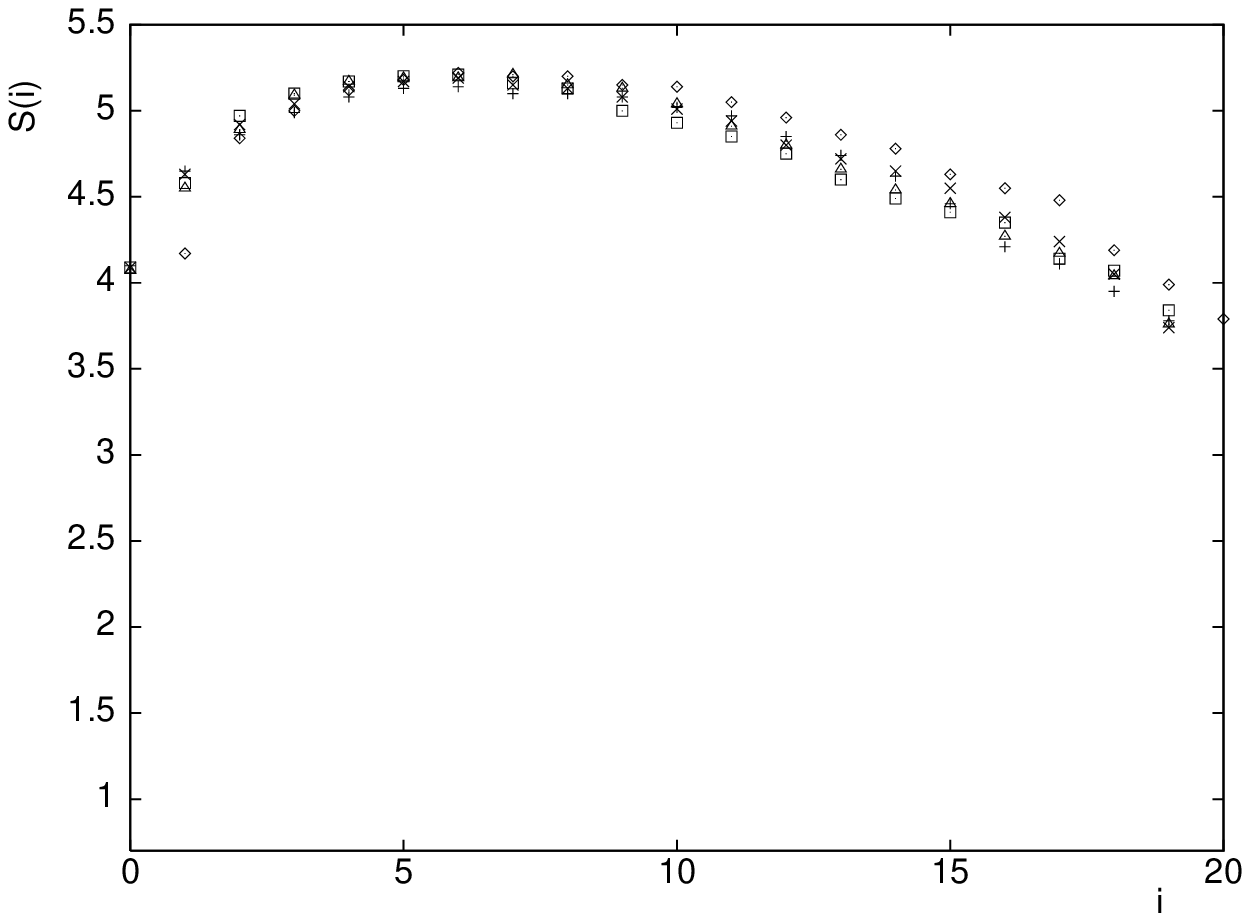}}
{\small
\vspace{0.4cm}
\begin{tabular}{cc}
~~~~~Fig.1a) Number of clusters for $i=1 \div 20$ &  ~~~  Fig.1b) Entropy for  5 SERENE events.\\
~~($\epsilon_i = 0.002*i$) for 5 SERENE events.&  \\
~~ Each  symbol corresponds to one  event. &\\
\end{tabular}}
\vspace{-1cm} 
\end{figure}
\begin{figure}
\centerline{%
\epsfxsize=7cm
\epsfbox{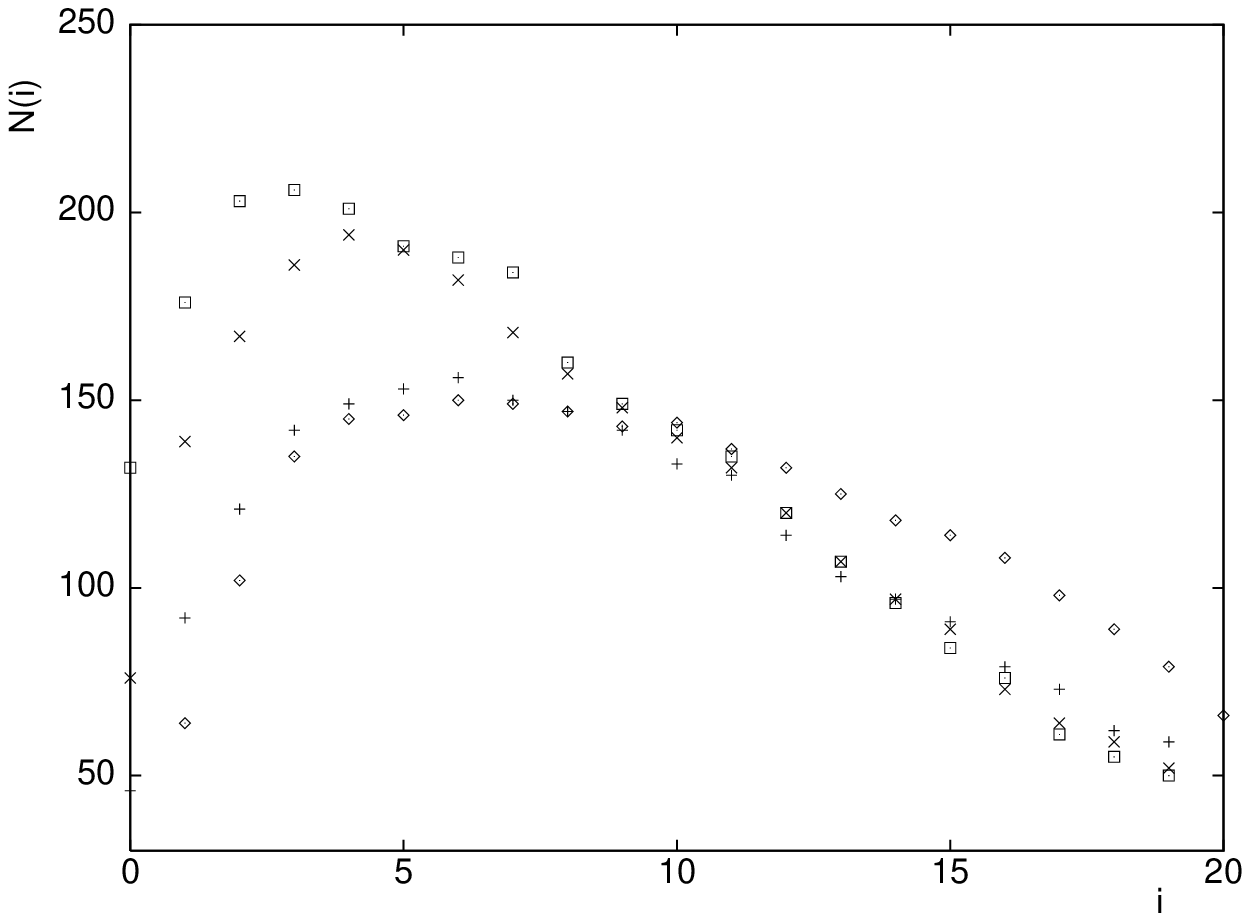}
\epsfxsize=7cm
\epsfbox{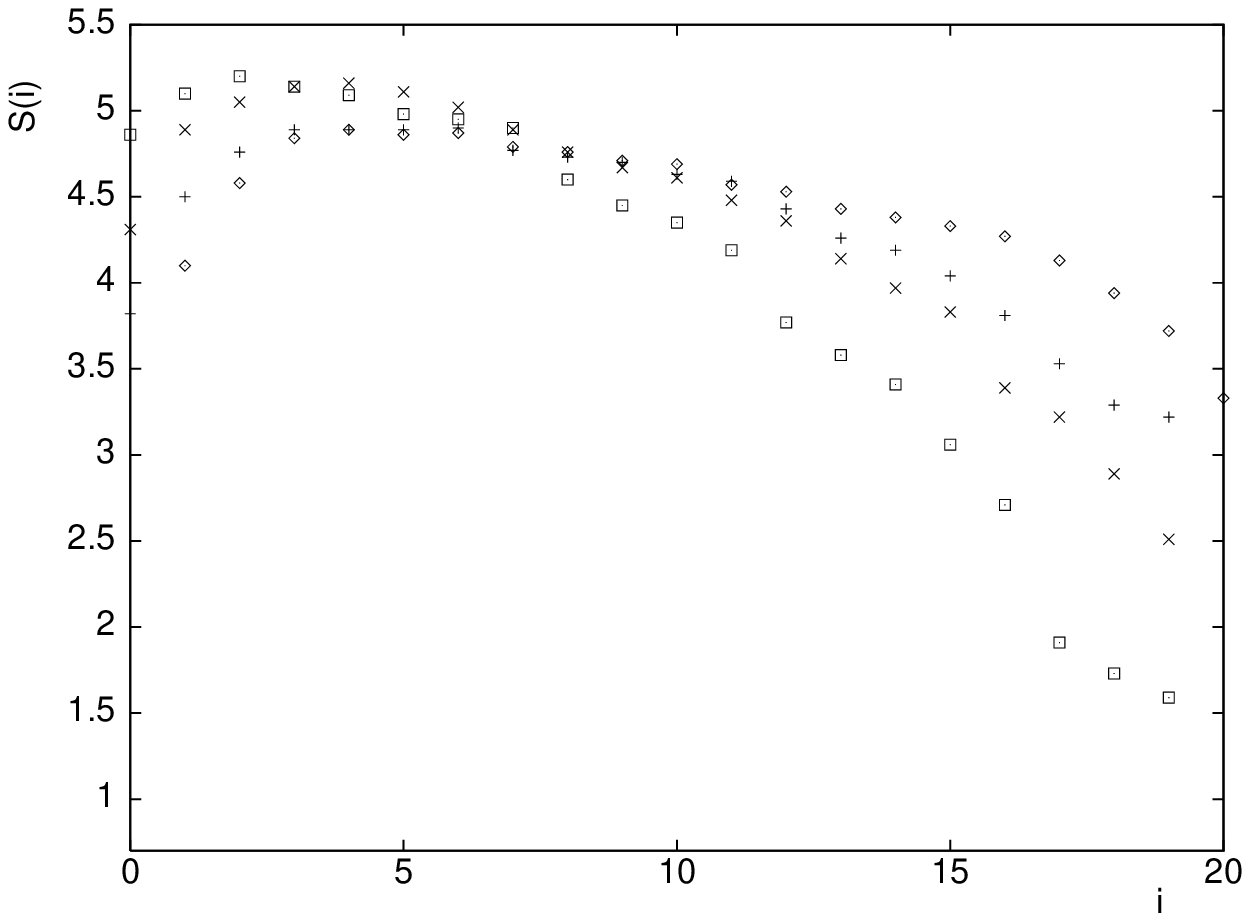}}
%\epsfig{file=exrkr.eps,width=7cm}
%\epsfig{file=ex1.eps,width=7cm}}
%\caption
{\small
\vspace{0.4cm}
\begin{tabular}{cc}
~~~~~~~~~~Fig.2a) Number of clusters for 4  &~~~~~~~~~~Fig.2b)  Entropy for 4 experimental \\
experimental events. &    events.
\end{tabular}} 
\vspace{-1cm}
\end{figure}
\begin{figure}
\centerline{%
\epsfxsize=7cm
\epsfbox{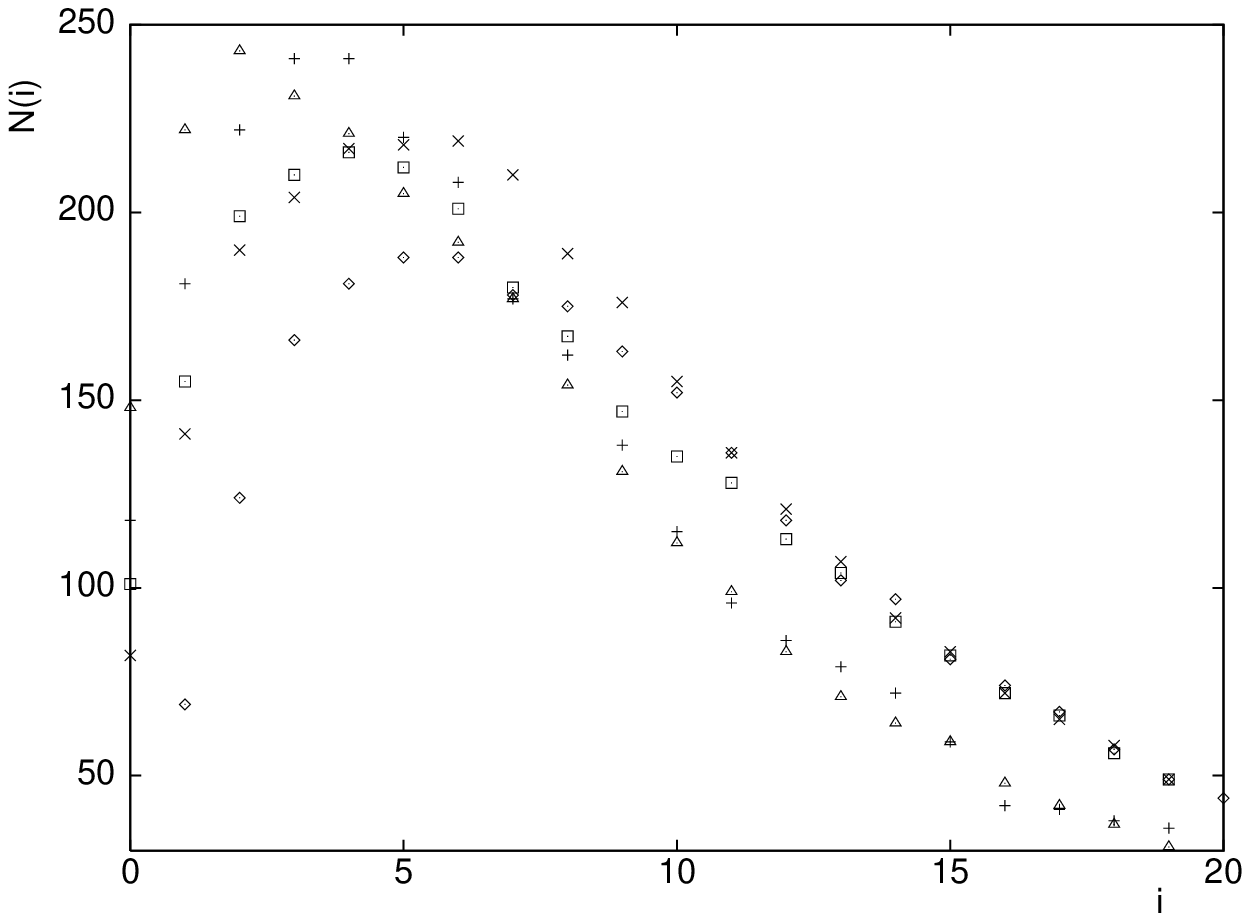}
\epsfxsize=7cm
\epsfbox{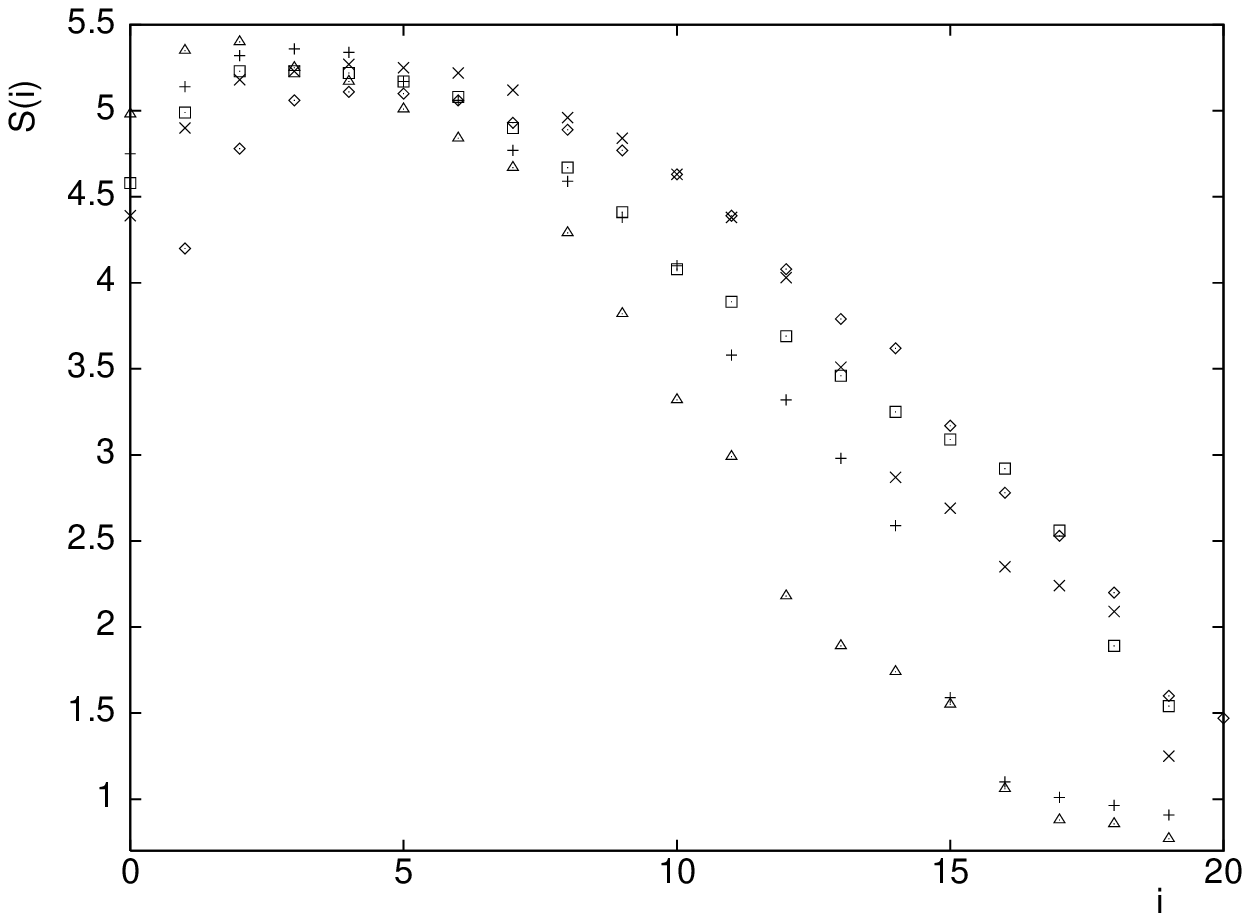}}
{\small
\vspace{0.4cm}
\begin{tabular}{cc}
~~~~~~~~~Fig.3a) Number of clusters for 5 VENUS  & Fig.3b) Entropy for 5 VENUS events.\\
events. &  \\
\end{tabular}}
\vspace{-1cm}
\end{figure}

\newpage
TAB.1. Values of $\rho_c$ and $\rho_s$. Number of particles in a cluster $>$ 1.\\
{\begin{center}
{\large 
\begin{tabular}{||c|c|c|c|c|c|c}
\hline
\hline
\multicolumn
{2}{||c|}{SERENE}&
\multicolumn {2}{c|}
{Exper.}&
\multicolumn {2}{c||}{VENUS}\\
\hline
\hline
   $\rho_c$ & $\rho_s$ &   $\rho_c$ & $\rho_s$ & $\rho_c$ & $\rho_s$\\
\hline
\hline
2.29& 1.37 &2.27 &1.47    & 4.27 &3.47      \\
\hline
2.32 &1.36 &
2.64 &1.52  & 6.69 &5.90    \\
\hline
2.35 &1.36 & 4.12 &3.27 &4.40 &3.40 \\
  \hline
2.34 &1.38 & 3.73 &2.06&    4.47 &4.12     \\
\hline
2.69 &1.39 & &  &7.83 &7.01     \\

\hline
\hline
\end{tabular}
}
\end{center}
}
\vspace{1cm}

TAB.2. Values of $\rho_c$ and $\rho_s$. Number of particles in a cluster $>$ 2. \\
{\begin{center}
{\large 
\begin{tabular}{||c|c|c|c|c|c||}
\hline
\hline
\multicolumn
{2}{||c|}{SERENE}&
\multicolumn {2}{c|}
{Exper.}&
\multicolumn {2}{c||}{VENUS}\\
\hline
\hline
$\rho_c$ & $\rho_s$ & $\rho_c$ & $\rho_s$& $\rho_c$ & $\rho_s$\\
\hline
\hline
1.63& 1.27 &1.66 &1.33    & 3.20 &3.46      \\
\hline
1.73 &1.26 & 1.86 &1.37  & 4.53 & 5.94    \\
\hline
1.63 & 1.25 & 3.48 &3.26 &3.52 &3.40 \\
  \hline
1.65 &1.27 & 2.82 & 1.93 &    3.81 &4.50     \\
\hline
1.95 &1.28 & &  &7.00 &7.54     \\

\hline \hline
\end{tabular}
}
\end{center}
}
\vspace{0.5cm}
\section{Conclusions}
Comparing  the results presented in Fig.1 and Fig.2 and the numbers given 
in Tab.1 and Tab.2 we may conclude that:
\begin{itemize}
\item all the SERENE events follow one ("quiet") pattern, where the fall of the number of 
clusters and entropy from its maximal value is rather slow; the broad entropy distribution 
reflects the randomness of these events' structure 
\item on the contrary, all the VENUS events exhibit a completely different 
behaviour; the majority of particles goes into one or a few clusters for larger values of 
$\epsilon$, which results in a fast fall of the number of clusters and entropy
\item two of the experimental events (the same as in Ref.~\cite{klm} and 
Ref.~\cite{fw}) are SERENE like, the other two resemble more the VENUS like pattern.
\end{itemize} 
\par 
Thus, using a slightly different language, we confirmed our previous results ~\cite{fw}. 
The events 
from data seem to be more differentiated than those from the generators.
Data with full momentum measurements and particle identification would be of great help 
in further analysis, improving our understanding of multiple production
in heavy ion collisions.

\section{Acknowledgements}
We used the data kindly provided to us by B. Wosiek (Ref. ~\cite{klm})
and the data prepared from the "VENUS" generator. An interesting conversation with P. Bogu\'s 
concerning the application of the J-S entropy in medical data bases should be noticed. The financial support of KBN 
grants \# 2 P03B 086 14 and \# 2P03B 010 15 is gratefully acknowledged.  One 
of us (RW) is grateful for a partial financial support by the KBN grant \# 2 P03B 019 17.

\end{document}